\documentclass[twocolumn,preprintnumbers,amsmath,amssymb,showpacs]{revtex4}

\usepackage{graphicx}% Include figure files
\usepackage{dcolumn}% Align table columns on decimal point
\usepackage{bm}% bold math

\def\figin{\includegraphics[scale=0.47]}

\begin{document}

%\preprint{version 14 (\today)}

\title{Electron Spectra for $D$ and $B$ Meson Semi-leptonic Decays at RHIC from PYTHIA with
Modified Heavy Quark Fragmentation}% Force line breaks with \\

%\author{Xiaoyan Lin$^{a}$, An Tai$^{b}$, Huan Z. Huang$^{b}$, Lianshou Liu$^{a}$\\
%$^{a}$Institute of Particle Physics, Huazhong Normal University, Wuhan, China\\
%$^{b}$University of California, Los Angele}

\author{Xiaoyan Lin\\Institute of Particle Physics, Huazhong Normal University, Wuhan 430079, China\\
Department of Physics and Astronomy, University of California, Los Angeles, CA 90095, USA}

\date{\today}

\begin{abstract}
We present a study of charmed hadron transverse momentum
distribution from p+p collisions at $\sqrt{s_{NN}}=200$ GeV using
the PYTHIA Monte Carlo event generator. 
The PHENIX tuned PYTHIA parameter set based on electron 
measurements alone yields a $D$ meson
$p_{T}$ distribution softer than the preliminary
STAR experimental data from d+Au collisions scaled by the number of binary
collisions. In order to match the STAR $p_{T}$ spectrum, the higher order QCD effects
in PYTHIA have to be taken into account, and a fragmentation function of charm quarks
much harder than the default Peterson function is needed.
Electrons from beauty quark semi-leptonic decays are found significantly below the
contribution from charm quark decays for the $p_{T}$ region up to 8 GeV/c in
our modified fragmentation scheme, while in the default fragmentation scheme the $B$ decay
electrons dominate at $p_{T}$ above 4 GeV/c. We propose an experimental measurement
of electron and charged hadron correlation to differentiate these scenarios.
\end{abstract}


\maketitle

Heavy quarks are believed to be produced through initial parton-parton, mostly gluon-gluon,
scatterings in nuclear collisions at the Relativistic Heavy Ion Collider (RHIC) energies. 
Theoretical calculations of heavy quark production within the perturbative Quantum ChromoDynamics (pQCD)
framework is considered more reliable because the heavy quark mass sets a natural scale for the pQCD.  
Charm quark production is sensitive to the incoming parton flux for the initial
conditions of nuclear collisions~\cite{appel}\cite{mull}.
The transport dynamics of the heavy quarks in nuclear
medium such as flow~\cite{rapp} and energy loss~\cite{doks}\cite{bw-zhang} can 
probe QCD properties of the dense matter created in nucleus-nucleus collisions. Therefore, heavy
quark measurements provide unique insights into QCD properties of the new state of matter produced 
in nucleus-nucleus collisions.

The heavy quark production in p+p or p+A collisions provides a reference for heavy meson
formation and for nuclear modification factors of heavy quarks in nuclear medium. 
The charm fragmentation function has been measured in $e^+e^-$ and
$\gamma p$ collisions and was fit by the Peterson function~\cite{peterson}
(defined later in the paper) with a parameter $\epsilon \approx 0.05$ 
(default in PYTHIA~\cite{pythia})~\cite{cfrag}. However, in charm hadroproduction, it was
observed that the c-quark $p_{T}$ distributions of 
next-to-leading-order (NLO) pQCD calculations agree well with the
measured open charm $p_{T}$ spectrum~\cite{barequark}, 
indicating that a much harder fragmentation function peaked at 
$z \approx 1$ is needed in charm hadroproduction. A more detailed
discussion of this observation can be found in~\cite{cbigpaper}.
Coalesence~\cite{ZW-lin} or recombination~\cite{duke,hwa} models have also been proposed
for charmed meson formation by combining a charm quark with 
a light up or down quark, presumably with
soft $p_T$~\cite{coal}. Thus the charmed hadron $p_T$ would coincide with the bare charm quark
$p_T$ distribution in this hadronization scheme. These various hadron formation schemes can lead to
significantly different interpretation of electron $p_T$ spectra from experimental measurements.

Recently the PHENIX collaboration reported a measurement of non-photonic
electrons from heavy quark semi-leptonic decays from $pp$ and $AA$ 
collisions~\cite{phenix1}\cite{phenix2}. The measured electron yields
appear to follow approximately a binary scaling. Such a scaling would
imply that the charm quarks have a small energy loss in the dense medium
created at RHIC. However, such an interpretation is subject to the ambiguity
that the electron contribution from $B$ quarks, which is predicted to have
a much smaller energy loss than that of charm quarks~\cite{magdalena}, is unknown experimentally
in the $p_T$ region above 3-4 GeV/c. In order to evaluate heavy quark energy
loss using non-photonic electron measurement, contributions from
$D$ and $B$ meson decays have to be investigated.

In this paper we evaluate the $p_{T}$ distribution of $D$ mesons
from PYTHIA v6.22~\cite{pythia} and compare the PYTHIA results
with the STAR preliminary measurements~\cite{an}. The charm quark
fragmentation function will be modified from the default Peterson
function and the PYTHIA parameters are tuned in order 
to match the PYTHIA results on $p_{T}$ distribution with the
experimental data. The electrons from $D$ meson decays are
compared with STAR preliminary electron $p_{T}$ distribution~\cite{suaide}. We
present a method to qualitatively estimate the significance of $B$
meson decays to the electron spectrum at high $p_{T}$ and to
verify the scheme of modified heavy quark fragmentation function as an
explanation for hard $p_{T}$ distribution of $D$ meson in $pp$ collisions at RHIC.

In our PYTHIA calculations, we started with a PHENIX tuned
parameter set~\cite{phenix1}, and changed the value of PARP(67) to 4, which
enhances c-quark production probability through gluon
splitting~\cite{lundc}. In addition, we also modify the parameter
in the Peterson function so that PYTHIA calculation approximately follows the shape of
the STAR measured open charm $p_{T}$ spectrum.

\begin{figure}
\figin{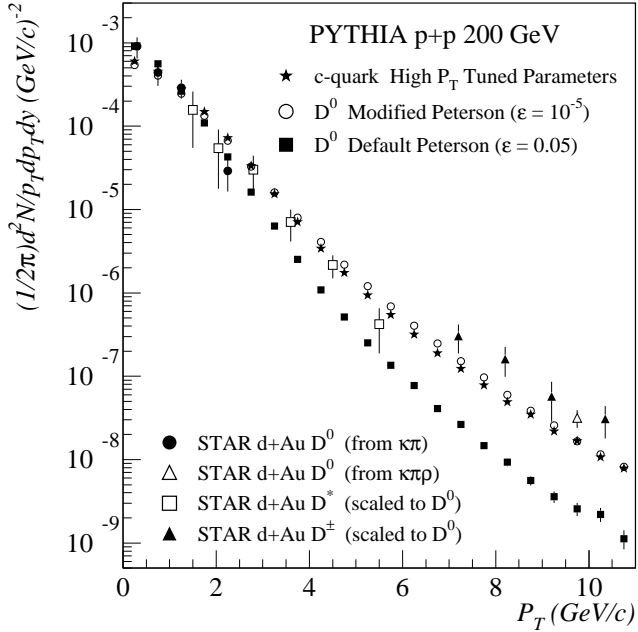}
\caption{\label{fig:D0}
$D^0$ and c-quark spectra from PYTHIA calculations compared with STAR preliminary
data from d+Au collisions scaled by $N_{bin}=7.5$.}
\end{figure}

Fig.~\ref{fig:D0} shows our PYTHIA calculations together with the STAR open charm
spectrum~\cite{an},
where the measured open charm data points from 
d+Au collisions at $\sqrt{s_{NN}}=200$ GeV
have been divided by $N_{bin}=7.5$ -- the number of
binary collisions.
The PYTHIA spectra have been normalized to the measured 
$dN/dy$ for $D^0$ at mid-rapidity ($0.0265\pm 0.0036$({\it
stat.})$\pm 0.0071$({\it syst.})).

The charm quark $p_{T}$ distribution labeled by stars in
Fig.~\ref{fig:D0} is from PYTHIA calculation with PHENIX tuned parameters 
and PARP(67)=4. We will refer this
set of parameters as high-$p_{T}$-tuned parameters in the rest of the paper. 
The PARP(67)=4 was introduced to take into account higher order effects in pQCD
calculation. We found that the change mainly affects c-quark production 
at high $p_{T}$. Due
to the large uncertainties in scaling $D^{\star}$ and $D^{\pm}$ to
$D^0$ ($D^{*}/D^{0}\approx D^{\pm}/D^{0}=0.4\pm 0.09 \pm 0.13 $), the STAR 
measurement of $D$ $p_T$ shape has large systematic uncertainties in the high $p_T$ region.
Without fine tuning any other PYTHIA parameters, we found that the
bare charm quark spectrum approximately match the STAR open charm $p_T$ distribution.
This also indicates that PARP(67)=4 somehow does generate a $p_T$ distrbution
similar to the actual NLO pQCD calculation~\cite{vogt}, which coincides with the
STAR $p_T$ spectrum as well~\cite{an}.  
It is not the purpose of this paper to show that the
PYTHIA calculation with our parameter set and NLO pQCD calculation are equivalent. 
The PYTHIA calculation with the high-$p_{T}$-tuned parameter set seems to reproduce 
the same observation seen when
comparing NLO pQCD calculation for c-quark to measured open charm
data~\cite{cbigpaper}. 
We will further modify the fragmentation function in PYHTIA
and investigate the consequences on single electron measurements.

\begin{figure}
\figin{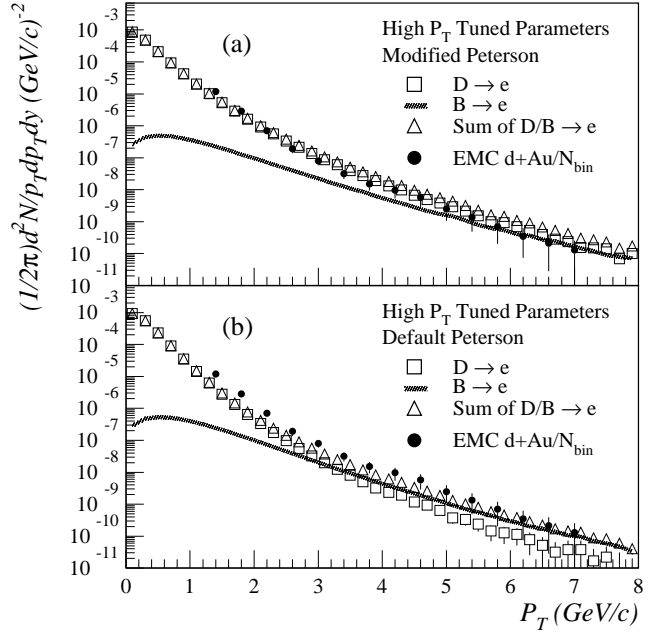}
\caption{\label{fig:elec}
(a) Electron spectra from PYTHIA calculation with the high-$p_{T}$-tuned parameters
and modified Peterson function for charm quarks and beauty quarks compared with background
subtracted single electron spectrum measured by STAR EMC from d+Au collisions scaled
by $N_{bin}=7.5$. (b) Electron spectra from PYTHIA calculation with the
high-$p_{T}$-tuned parameters and default Peterson function for charm quarks and beauty quarks.}
\end{figure}

The solid squares in Fig.~\ref{fig:D0} depict $D^0$ $p_{T}$ distribution from PYTHIA
calculation with the high-$p_{T}$-tuned parameters and the default Peterson fragmentation
function in PYTHIA. The default Peterson function refers to the value
of the parameter ${\varepsilon}$ in Peterson function
\begin{eqnarray}
D(z {\equiv} p_{D}/p_{c}) {\propto} \frac {1}{z[1-1/z-{\varepsilon}/(1-z)]^2}
\end{eqnarray}
being 0.05 for charm quarks and 0.005 for beauty quarks. The
default Peterson fragmentation for charm quarks is too soft to
reproduce the measured open charm spectrum. We modified the value
of the parameter ${\varepsilon}$ to make the fragmentation
function harder. The result is shown as open circles in
Fig.~\ref{fig:D0}. The $D^0$ spectrum roughly coincides with the
measured data. Here the value of ${\varepsilon}$ we used is
$10^{-5}$ for both charm quarks and beauty quarks. In this case
the fragmentation function is very like $\delta(1-z)$. Hence, the
charm quark hardly loses its momentum during fragmentation. The
PYTHIA calculation with high-$p_{T}$-tuned parameter set and modified
Peterson fragmentation function (${\varepsilon}=10^{-5}$) can
reasonably depict the measured open charm data. 
A harder fragmentation function is needed for the hadronization of
charm quarks if the pQCD calculation is to describe the STAR
preliminary data.

\begin{figure}
\figin{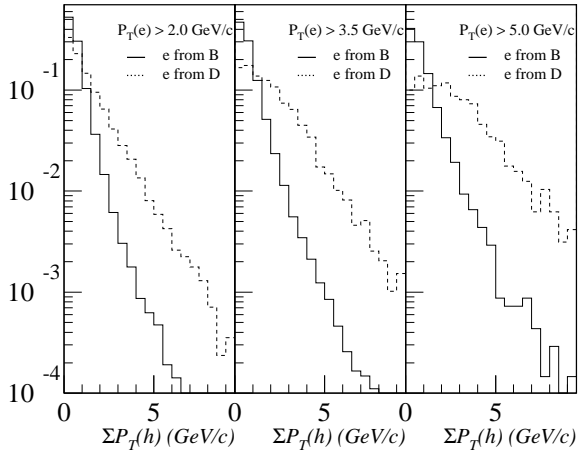}
\caption{\label{fig:sumpt}
Sum $p_{\rm T}$ distributions for charged hadrons in the electron cone. Solid lines show
electrons from $B$ meson decay case. Dashed lines show electrons from $D$ meson decay case.
Left panel: Electrons selected at $p_{T} > 2.0$ GeV/c. Middle panel: Electrons selected
at $p_{T} > 3.5$ GeV/c. Right panel: Electrons selected at $p_{T} > 5.0$ GeV/c.}
\end{figure}

\begin{figure}
\figin{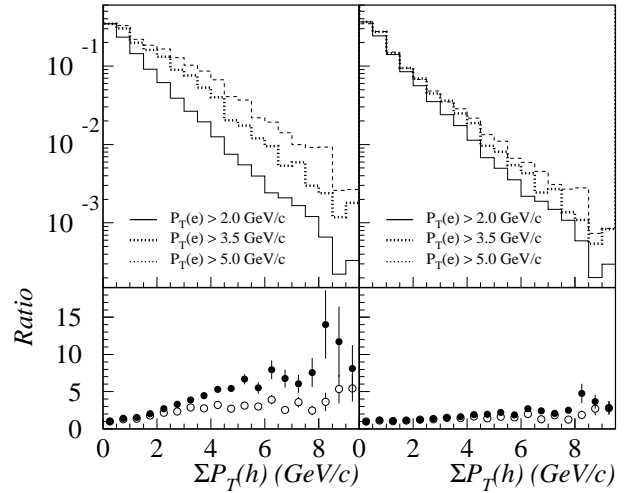}
\vskip -1.7cm
\caption{\label{fig:mixedsumpt}
Upper panels: Charged hadron sum $p_{\rm T}$ distributions of $D$ decay case and $B$ decay
case weighted by the fractions of electrons from $D$ and $B$. The case that $D$ meson decay
contribution is always dominant is shown in the left panel. The case that $B$ meson decay
contribution become dominant at $p_{T} > 4.0$ GeV/c range is shown in the right panel.
Bottom panels: The ratios of $p_{T} > 5.0$ GeV/c case to $p_{T} > 2.0$ GeV/c case are
shown as solid circles. The ratios of $p_{T} > 3.5$ GeV/c case to $p_{T} > 2.0$ GeV/c
case are shown as open circles.}
\end{figure}

The STAR independent measurements of the reconstructed $D^0$ and
single electrons from heavy quark semileptonic decays measured
with TOF and TPC are consistent~\cite{adams}. The electron
measurement there only covers up to $p_{T} < 4$ GeV/c and has no
sensitivity to the $B$ meson contribution. We also checked the
consistency between open charm data and non-photonic single
electron data in d+Au collisions measured by STAR EMC~\cite{suaide} within our
PYTHIA calculation. The results are shown in Fig.~\ref{fig:elec}.
The PYTHIA spectra of electrons  from charm meson decays are
scaled by the same factor used to scale the PYTHIA $D^0$ spectra
to the measured $dN/dy(D^0)$ at mid-rapidity. The electron spectra
from beauty meson decay are normalized by the ratio of
%%                   ^^^
$\sigma_{b\bar b}$/$\sigma_{c\bar c}$ based on %%%
the NLO pQCD calculation~\cite{vogt}. The band corresponds to the
%%                    ^^^
theoretical uncertainty of this ratio, which is in the region of
0.45\% - 0.60\%~\cite{vogt}. We used the value at the center of
this range to calculate the sum of the electrons from both charm
and beauty meson decays. The EMC electron data points are divided
by $N_{bin}$. The electron spectra from the PYTHIA calculation
with the high-$p_{T}$-tuned parameters and modified Peterson
function for charm quarks and beauty quarks are shown in the upper
panel of Fig.~\ref{fig:elec}. 
The PYTHIA calculation with the high-$p_{T}$-tuned parameter set and modified
heavy quark fragmentation function can simultaneously describe the STAR 
direct open charm $p_T$ distribution and the STAR EMC electron data reported 
at the Quark Matter 2004 conference.
Our calculation indicates that the contribution of electrons from $B$
meson decays is not dominant for the measured $p_T$ region up to 8 GeV/c.

The measurement of electron $p_{T}$ distribution alone has a 
reduced sensitivity to the $p_{T}$ distribution of $D$ mesons as shown in reference~\cite{batsouli}. 
Within the large statistical
and systematic errors of the electron data, it is possible to use high-$p_{T}$-tuned
PYTHIA calculation and the default Peterson function for both $c$ and $b$ quarks to yield
an electron $p_{T}$ distribution similar to the measurement at high $p_{T}$. In this case,
as shown in the bottom panel of Fig.~\ref{fig:elec}, electrons from $b$ quark decays are
the dominant source for $p_{T}$ greater than 4 GeV/c.

In order to differentiate these two scenarios, we studied the
correlation between charm/beauty semileptonic decay electrons and
charged hadrons in PYTHIA with these two sets of parameter in
Peterson function. First, we select an electron decayed from $B$
or $D$ mesons at a given electron $p_{T}$ cut, and make a cone
around the direction of the electron. %%%
Then we study the scalar summed $p_{T}$ distributions of these
charged hadrons in the cone ($p_{T}$ refers to the transverse
momentum in the laboratory frame). Here the cone is defined by
$|\eta_{h} - \eta_{e}| < 0.35$ and $|\varphi_{h} - \varphi_{e}| <
0.35$ ($\eta$ is pseudorapidity and $\varphi$ is azimuthal angle).
The summed $p_{T}$ distributions of charged hadrons at three
different electron $p_{T}$ cuts are shown in Fig.~\ref{fig:sumpt}.
The distributions are normalized to unity. The solid lines in
Fig.~\ref{fig:sumpt} are 
electrons from $B$ meson decays and the dashed lines are
electrons from $D$
meson decays.
We found that the summed $p_{T}$                         
distributions do not change significantly between the default
Peterson function and our modified fragmentation function. When
electron $p_{T}$ cut goes from 2.0 GeV/c to 5.0 Gev/c, the summed
$p_{T}$ distributions broaden for both $B$ meson decay case and
$D$ meson decay case. However, the change is more drastic for the
$D$ meson decay case.

\begin{center}
\begin{table}
\caption{Fractions of $B$ meson decay contribution at three electron $p_{T}$ cuts from
two different PYTHIA calculations with high-$p_{T}$-tuned parameters. Modified Peterson
function and default Peterson function are each used for both charm quarks and
beauty quarks.}
\begin{tabular}{cc|c|c|c}
\hline
\hline
\multicolumn{2}{c|}{$p_{T}$ ( GeV/c )} &2.0 &3.5 &5.0  \\
\hline
& {Modified Peterson function} &7$\%$ & 22$\%$ & 32$\%$ \\
& {Default Peterson function} &17$\%$ & 54$\%$ & 69$\%$ \\
\hline
\hline
\end{tabular}
\label{tab:fraction}
\end{table}
\end{center}
\vskip -1.035cm

The upper panel of Fig.~\ref{fig:mixedsumpt} shows the sum of $D$
decay case and $B$ decay case weighted by fractions of electrons
from $D$ and $B$. Here the distributions are normalized such that
the values in the first bin are the same. In the left upper panel,
the fractions come from PYTHIA calculations with the
high-$p_{T}$-tuned parameters and modified Peterson function for
charm quarks and beauty quarks, while the fractions in the right
upper panel of Fig.~\ref{fig:mixedsumpt} are from PYTHIA
calculations with the high-$p_{T}$-tuned parameters, but default
Peterson function for charm quarks and beauty quarks. The values
of the fractions are shown in Table I. When electron $p_{T}$
increases, if $B$ meson decays become dominant, the summed $p_{T}$
distribution of charged hadrons changes slightly as shown in the
right upper panel of Fig.~\ref{fig:mixedsumpt}. However, if the
$D$ meson decay contribution is always dominant, the summed $p_{T}$
distribution broadens drastically as shown in the left upper panel
of Fig.~\ref{fig:mixedsumpt}. 
Ratios are shown in the bottom panels in
Fig.~\ref{fig:mixedsumpt}. The solid cirles are the ratios of
$p_{T} > 5.0$ GeV/c case over $p_{T} > 2.0$ GeV/c case. The open
cirles are the ratios of $p_{T} > 3.5$ GeV/c case over $p_{T} >
2.0$ GeV/c case. 
Experimental observation of these ratios can be used to determine the
relative significance of $B$ and $D$ decay electrons.

In conclusion, we found that in order to match the STAR measurement of $p_{T}$ shape of
$D$ mesons from d+Au collisions at RHIC using PYTHIA Monte Carlo generator, higher order
QCD effects have to be taken into account (PARP(64)=4) and a harder charm quark fragmentation
function must be used. The high-$p_{T}$-tuned PYTHIA calculation with the modified
fragmentation function can simultaneously describe the $p_{T}$ distributions of $D$ mesons
and non-photonic electrons from semi-leptonic decays of heavy quarks. Because the measurement of 
electrons alone
has a reduced sensitivity on the $p_{T}$ shape of $D$ mesons, the high-$p_{T}$-tuned
PYTHIA calculation with default Peterson fragmentation function for both $c$ and $b$ quarks can
match the STAR electron measurement at high $p_{T}$ within large statistical and systematic
errors. This calculation predicts that the $B$ decayed electrons dominate at the high $p_{T}$
($>$ 4 GeV/c) region. However, in the proposed scheme of modified charm and beauty quark
fragmentation, the beauty quark decay electrons are not a dominant contribution over
the entire $p_{T}$ region up to 8 GeV/c. Using PYTHIA as a guidance we found that the
summed $p_{T}$ of charged hadrons within a narrow cone of heavy quark decayed electrons can be
used to differentiate these scenarios.

\begin{acknowledgements}
X.Y. Lin wishes to thank An Tai, Huan Z. Huang and Lianshou Liu for their helpful 
discussions. This work was supported in part by NSFC under project 10375025.
\end{acknowledgements}

{}

\begin{thebibliography}{}
\bibitem{appel} J.A. Appel, Annu. Rev. Nucl. Part. Sci. $\bf{42}$, 367 (1992).
\bibitem{mull} B. M\"{u}ller and X.N. Wang, Phys. Rev. Lett. $\bf{68}$, 2437 (1992).
\bibitem{rapp} V. Greco, C.M. Ko and R. Rapp, Phys. Lett. B {\bf 595}, 202 (2004).
\bibitem{doks} Y.L. Dokshitzer and D.E. Kharzeev, Phys. Lett. B $\bf{519}$, 199 (2001).
\bibitem{bw-zhang} Ben-Wei Zhang {\it et at}., Phys. Rev. Lett. $\bf{93}$, 072301 (2004).
\bibitem{peterson} C. Peterson {\it et al}., Phys. Rev. D {\bf 27}, 105 (1983).
\bibitem{pythia} T. Sj\"ostrand, Comput. Phy. Commun. $\bf{135}$, 238 (2001).
\bibitem{cfrag} P. Nason and C. Oleari, Nucl. Phys. B $\bf{565}$, 245 (2000);
L. Gladilin (for ZEUS Collaboration), hep-ex/0309044.
\bibitem{barequark} M. Adamovich {\sl et al}., Beatrice Collaboration, Nucl.
Phys. B $\bf{495}$, 3 (1997); G.A. Alves {\sl et al}., E769 Collaboration, Phys. Rev.
Lett. $\bf{77}$, 2392 (1996).
\bibitem{cbigpaper} S. Frixione {\sl et al}., Adv. Ser. Direct. High Energy
Phys. $\bf{15}$, 609 (1998).
%\bibitem{pythia} T. Sj\"ostrand, Comput. Phy. Commun. $\bf{135}$, 238 (2001).
\bibitem{ZW-lin} Z.W. Lin and C.M. Ko, Phys. Rev. Lett. $\bf{89}$, 202302 (2002).
\bibitem{duke} R.J. Fries {\sl et al}., Phys. Rev. Lett. $\bf{90}$, 202303 (2003).
\bibitem{hwa} R.C. Hwa and C.B. Yang, Phys. Rev. C $\bf{67}$, 034902 (2003).
\bibitem{coal} R. Rapp and E.V. Shuryak, Phys. Rev. D $\bf{67}$ 074036 (2003).
\bibitem{phenix1} K. Adcox {\sl et al}., PHENIX Collaboration, Phys. Rev. Lett. $\bf{88}$,
192303 (2002).
\bibitem{phenix2} S.S. Adler {\sl et al}., PHENIX Collaboration, nucl-ex/0409028.
\bibitem{magdalena} M. Djordjevic and M. Gyulassy, J. Phys. G $\bf{30}$, S1183 (2004).
\bibitem{an} An Tai (for the STAR Collaboration), J. Phys. G: Nucl. Part. Phys.
$\bf{30}$, S809-S817 (2004).
\bibitem{suaide} A.A.P. Suaide (for the STAR Collaboration), J. Phys. G: Nucl. Part. 
Phys. $\bf{30}$, S1179-S1182 (2004).
\bibitem{lundc} E. Norrbin and T. Sj\"ostrand, Eur. Phys. J. C $\bf{17}$, 137 (2000).
\bibitem{vogt} Vogt R., hep-ph/0203151.
\bibitem{adams} J. Adams {\sl et al}., STAR Collaboration, nucl-ex/0407006.
\bibitem{batsouli} S. Batsouli {\sl et al}., Phys. Lett. B $\bf{557}$, 26 (2003).

\end{thebibliography}
\end{document}